\documentclass[12pt,a4paper]{article}
\usepackage{amsmath,amssymb}
\usepackage[dvipdfmx]{graphicx}
 
\newcommand{\GeV}{\rm{GeV}}
\newcommand{\MeV}{\rm{MeV}}

\newcommand{\lrf}[2]{ \left(\frac{#1}{#2}\right)}
\newcommand{\lrfp}[3]{ \left(\frac{#1}{#2} \right)^{#3}}

\def\Ar#1{A_{\rm r}({#1})}
\def\Eb#1{E_{\rm b}({#1})}

\def\Rb87{{}^{87}{\rm Rb}}
\def\C12{{}^{12}{\rm C}}
\def\Kr86{{}^{86}{\rm Kr}}
\def\oo16{{}^{16}{\rm O}}

\begin{document}

\begin{titlepage}

\begin{center}

\hfill UT-12-31\\
\hfill IPMU12-172\\

{\Large \bf 

Constraints on Hidden Photon Models\\ 
from Electron $g-2$ and Hydrogen Spectroscopy

}

\vskip .75in

{\large
Motoi Endo$^{(a,b)}$, 
Koichi Hamaguchi$^{(a,b)}$, 
Go Mishima$^{(a)}$
}

\vskip 0.25in

{\em
$^a$ Department of Physics, The University of Tokyo,
Tokyo 113-0033, Japan\\
$^b$Kavli Institute for the Physics and Mathematics of the Universe, 
The University of Tokyo, Kashiwa 277-8568, Japan
}

\end{center}
\vskip .5in

\begin{abstract}
The hidden photon model is one of the simplest models which can explain the anomaly of the muon anomalous magnetic moment ($g-2$). The experimental constraints are studied in detail, which come from the electron $g-2$ and the hydrogen transition frequencies. The input parameters are set carefully in order to take dark photon contributions into account and to prevent the analysis from being self--inconsistent. It is shown that the new analysis  provides a constraint severer by more than one order of magnitude than the previous result. 
\end{abstract}

\end{titlepage}

\section{Introduction}

A lot of experimental efforts have been devoted to searches for physics beyond the standard model (SM). 
One of the evidences for the existence of the new physics has been found in the precise measurement of the anomalous magnetic moment of the muon (muon $g-2$). The experimental result of the muon $g-2$ reported by the Brookhaven E821 experiment~\cite{Bennett:2006fi} deviates from the SM prediction with the updates on the hadronic contributions~\cite{Hagiwara:2006jt}:
\begin{equation}
  a_\mu({\rm exp}) - a_\mu({\rm SM}) = (26.1 \pm 8.0) \times 10^{-10},
  \label{eq:muon_g-2}
\end{equation}
which corresponds to more than 3 sigma level deviation, where Ref.~\cite{Prades:2009tw} is referred to for the hadronic light-by-light contribution. It is obviously too early to conclude that this is the signal of the new physics. Nonetheless, it is interesting to investigate the new physics models which can explain the muon $g-2$ anomaly.

The current deviation of the muon $g-2$ is comparable to the electroweak contribution of the SM. Thus, 
the particles of the new physics should have large couplings to the muon if they are as heavy as $\mathcal{O}(100)\,\GeV$,
or they must be light enough to explain the anomaly if the  interactions with the SM are 
 suppressed to avoid experimental constraints. A simple model of the latter case is the hidden photon model~\cite{Fayet:2007ua,Pospelov:2008zw}. In presence of a U(1) symmetry in a hidden sector, the extra U(1) gauge boson, so-called the hidden photon, can couple to the SM particles through a kinetic mixing with the gauge boson of the SM hypercharge. The hidden photon contributes to the muon $g-2$ at the one-loop level similarly to the SM photon. This becomes effective if the hidden photon mass is $\mathcal{O}(1-100)\,\MeV$ and the mixing angle is $\mathcal{O}(10^{-(2-3)})$. 

It is important to examine whether the hidden photon model satisfies the other experimental constraints. The severest upper bound on the hidden photon mass which can explain the muon $g-2$ anomaly is provided by searches for the hidden photon $A'$ in the processes, $e^+e^- \to \phi \to A' (\to e^+e^-)\, \eta (\to \pi^+\pi^-\pi^0)$, at the KLOE experiment~\cite{Archilli:2011zc} and $e^- N\to e^- N A' (\to e^+ e^-)$ at the APEX experiment~\cite{Abrahamyan:2011gv}.
 On the other hand, the lower mass bounds have been obtained by the electron beam dump experiment at the Fermilab E774~\cite{Bross:1989mp} and the electron $g-2$~\cite{Pospelov:2008zw}. In particular, the hidden photon contributes to the electron $g-2$ at the one-loop level similarly to the muon $g-2$. Recently, the SM prediction of the electron $g-2$ has been greatly improved. The tenth-order calculation of the QED contribution was completed~\cite{Aoyama:2012wj}, and the measurement of the fine structure constant was improved
 by the update on the measurement of the rubidium mass~\cite{FineStructureConstant}. It has been found that the experimental result~\cite{Hanneke:2008tm} agrees with the SM prediction~\cite{Aoyama:2012wj}:
\begin{equation}
  a_e({\rm exp}) - a_e({\rm SM}) = -(1.06 \pm 0.82) \times 10^{-12},
  \label{eq:electonr_g-2}
\end{equation}
where the dominant uncertainties come from the experimental result of the electron $g-2$ and the determination of the fine structure constant. 
This constraint is improved by one order of magnitude after the analysis in Ref.~\cite{Pospelov:2008zw}. 
Thus, the electron $g-2$ can provide a tighter constraint on the hidden photon parameter especially in a light mass region.
In this paper, we will revisit the constraint on the hidden photon from the electron $g-2$.\footnote{
Contributions of new physics models to the electron $g-2$ are also studied in Ref.~\cite{Giudice:2012ms}.
}

In addition to the direct contribution to the electron $g-2$ at the one-loop level, the SM prediction of the electron $g-2$ can be subject to the effect of the hidden photon through the determination of the fine structure constant. The fine structure constant
which is determined independently of the electron $g-2$
 is currently set by a precise measurement of the rubidium mass. The present best result is known to be~\cite{FineStructureConstant}
\begin{equation}
  \alpha^{-1}({\rm Rb}) = 137.035\ 999\ 049\ (90)\quad [6.6\times 10^{-10}],
  \label{eq:alpha}
\end{equation}
by combining several atomic parameters, where the leading uncertainty stems from the rubidium mass. Since the precisions of each measurements are as good as or better than that of the electron $g-2$, it is crucial to evaluate the hidden photon contribution to the fine structure constant for the purpose of deriving the electron $g-2$ constraint. In this paper, we will study the measurement of the fine structure constant and its correction from the hidden photon in detail. 

Among the atomic parameters, transition frequencies of the hydrogen are one of the most precisely measured parameters. They are utilized to determine fundamental parameters such as the Rydberg constant~\cite{Mohr:2012tt}. The SM prediction of the transition frequencies has been studied well, which is compared to the experimental results at the extremely precise level. In the hidden U(1) model, the atomic energy levels are shifted by the hidden photon at the tree level. Thus, the hidden photon parameter can be constrained tightly by the transition frequencies, independently of the electron $g-2$~\cite{Jaeckel:2010xx,earlyworks}. In this paper, we revisit the hidden photon contribution to the transition frequencies in addition to the lepton $g-2$ and the fine structure constant. 
A special care is taken to ensure that the input parameters are determined independently of the target observables such as the transition frequencies and the electron $g-2$, in order to make the analysis self--consistent.


\section{Lepton $g-2$}

The Lagrangian terms of the hidden photon model, which are relevant to the contributions to the lepton $g-2$, are given by
\begin{equation}
  \mathcal{L}_V = -\frac{1}{4} F^V_{\mu\nu} F^{V\,\mu\nu} 
  + \frac{1}{2} m_V^2 V_\mu V^\mu 
  + \frac{1}{2} \frac{\epsilon}{\cos \theta_W} F^V_{\mu\nu} B^{\mu\nu},
  \label{eq:Lagrangian}
\end{equation}
where the field/index, $V$, represents the hidden photon, and $B_{\mu\nu}$ is the field strength of the SM hypercharge. The second term in the right-handed side is a mass of the hidden photon. We  assume that the hidden photon mass is induced by the Higgs mechanism in the hidden sector. The last term corresponds to the kinetic mixing between the hidden and SM U(1)'s, where the mixing $\epsilon$ is conventionally scaled by the Weinberg angle, $\cos \theta_W$. In a low energy scale, by canonicalizing the kinetic terms and diagonalizing the mass matrix, the effective Lagrangian of the hidden photon becomes 
\begin{equation}
  \mathcal{L}_V = -\frac{1}{4} F^{A'}_{\mu\nu} F^{A'\,\mu\nu} 
  + \frac{1}{2} m_V^2 A'_\mu A'^\mu 
  - e \epsilon A'_\mu J^\mu_{\rm em}
  + g \epsilon \tan\theta_W \frac{m_V^2}{m_Z^2} A'_\mu J^\mu_Z,
  \label{eq:effLagrangian}
\end{equation}
where $A'$ is the hidden photon in the mass eigenstate,  $J^\mu_{\rm em} = \bar \psi \gamma^\mu \psi$ is the electromagnetic current of the charged leptons, and $J^\mu_Z$ is the weak neutral current.

There are two types of the contributions to the lepton $g-2$ from the hidden photon. The first one is a radiative correction to the lepton--photon vertex, which is evaluated at the one-loop level as~\cite{Fayet:2007ua,Pospelov:2008zw} 
\begin{eqnarray}
  \Delta a_\ell^{A'{\rm loop}} 
  &=& \frac{\alpha_0}{2\pi} \epsilon^2
  \left[ 
  \int^1_0 dz\, \frac{2 z(1-z)^2 m_\ell^2}{z m_V^2 + (1-z)^2 m_\ell^2} 
  + \mathcal{O} \left( \frac{m_\ell^2}{m_V^2} \frac{m_V^2}{m_Z^2} \right) 
  \right]
  \label{eq:lepton_g-2_exact}
  \\
  &=& \frac{\alpha_0}{3\pi} \epsilon^2 \frac{m_\ell^2}{m_V^2} 
  \left[ 1 + \mathcal{O} \left( \frac{m_\ell^2}{m_V^2} \right) + \mathcal{O} \left( \frac{m_V^2}{m_Z^2} \right) \right],
  \label{eq:lepton_g-2}
\end{eqnarray}
where $\alpha_0 \equiv e^2/4\pi$ is the fine structure constant of the Lagrangian. The first term in Eq.~\eqref{eq:lepton_g-2_exact} is the one-loop contribution of the electromagnetic current $J^\mu_{\rm em}$, and $\mathcal{O}(m_V^2/m_Z^2)$ is from $J^\mu_Z$. In the last equation, $\mathcal{O}(m_\ell^2/m_V^2)$ appears by expanding the integrand of Eq.~\eqref{eq:lepton_g-2_exact}. 
When the lepton is the electron, the expansion is valid for $m_V \gtrsim \mathcal{O}(1)\,\MeV$, while the contribution to the muon $g-2$ needs the evaluation without the expansion for $m_V \sim m_\mu$. 

The second contribution to the lepton $g-2$ is through the determination of the fine structure constant. The one-loop radiative correction of the photon provides the leading contribution to the lepton $g-2$ as
\begin{equation}
  a_\ell({\rm QED;\,leading}) = \frac{\alpha_0}{2\pi}.
\end{equation}
The higher order corrections are suppressed by orders of $\alpha_0$. Here, the fine structure constant $\alpha_0$ is defined as a parameter of the Lagrangian. The determination of the numerical value of $\alpha_0$ is based on the experimental measurements of the atomic parameters. In the literature, it is hypothesized that the SM is correct in the process of the $\alpha_0$ determination, and the theoretical prediction of the electron $g-2$ is based on that SM hypothesis. However, it should be noted that this procedure can be affected by the hidden photon. Then, the fine structure constant in the SM hypothesis, $\alpha^{\rm SM-hyp.}$, does not coincide with the Lagrangian parameter $\alpha_0$, but is corrected by the hidden photon effect as $\alpha^{\rm SM-hyp.} = \alpha_0 + \delta \alpha^{A'}$. Consequently, the leading QED calculation of the photon radiative correction provides an extra contribution from the hidden photon apart from the SM prediction as
\begin{eqnarray}
  \Delta a_\ell^{{\rm from}\;\alpha} &=&  
a_\ell^{\rm QED} (\alpha_0) - a_\ell^{\rm QED}(\alpha^{\rm SM-hyp.})    
  \nonumber \\
&\simeq &  
-\frac{\delta \alpha^{A'}}{2\pi}.
\end{eqnarray}
In the next section, the determination of the fine structure constant will be investigated in detail.


\section{Fine structure constant}
\label{sec:alpha}
The fine structure constant, which is determined independently of the electron $g-2$ and used as an input for the theoretical calculation of the electron $g-2$, is determined via the 
Rydberg constant,
\begin{eqnarray}
R_{\infty} &\equiv& \frac{\alpha_0^2 m_e c}{2 h},
\label{eq:Rydberg}
\end{eqnarray}
where $\alpha_0$ is the theoretically given fine structure constant,
$m_e$ is the electron mass, 
$c$ is the speed of light, and
$h$ is the Planck constant.
The most precise inferred value of the fine structure constant (except for the electron $g-2$)  is determined by the following relation~\cite{FineStructureConstant}
\begin{eqnarray}
\alpha_{\rm exp}^2 &=& 
\lrf{2 R_{\infty}}{c}
\lrf{\Ar{\Rb87}}{\Ar{\rm e}}
\lrf{h}{m_{\rm Rb}},
\label{eq:alpha_exp}
\end{eqnarray}
namely, by the precise measurement of $R_{\infty}$, $\Ar{\Rb87}$, 
$\Ar{\rm e}$, and $m_{\rm Rb}/h$. Here, $\Ar{X} \equiv m_X / (m_{\C12} / 12)$ is the relative atomic mass (relative isotopic mass). In fact, the mass ratios are measured more precisely than their absolute values by experiments. On the other hand, the absolute mass has been measured very precisely for the rubidium atom. The progress of the mass measurement is the main part of the improvement of the recent result of the fine structure constant~\cite{FineStructureConstant}.

Within the SM, if these quantities are measured with an infinite precision, the right-hand side of Eq.~(\ref{eq:alpha_exp}) gives rise to the theoretical value $\alpha_0^2$ itself. However,  in the presence of the hidden photon, the measurements of these quantities, which are implicitly  based on the SM hypothesis, may lead to values which are different from their definitions:
\begin{eqnarray}
X^{\rm SM-hyp.} = X \left(1 + \delta_X (\alpha_0, \epsilon, m_V) \right),
\end{eqnarray}
where $X = R_{\infty}, \Ar{\Rb87}, \Ar{\rm e}$, and $m_{\rm Rb}/h$. Here, $X^{\rm SM-hyp.}$ denotes the value of $X$ which is obtained when the determination is (wrongly) based on the SM only even in the presence of the hidden photon.
In the following, we discuss effects of the hidden photon to these quantities in turn.

\subsection{$R_{\infty}$}
The Rydberg constant $R_{\infty}$ is determined by the precise measurements of the hydrogen transition frequencies,
$(E_a-E_b)/h$. The theoretical prediction of the atomic energy levels, which depend on the Rydberg constant, has been known well in the SM, and the transition frequencies have been measured very precisely. The detailed procedure is found in Ref.~\cite{Mohr:2012tt}. 
In the presence of hidden photon, the energy levels of the states, $E_a$, are represented by
\begin{eqnarray}
\frac{E_a}{h} =  - R_{\infty}c
\left( \frac{1}{n_a^2} + \delta_a^{\rm SM}(\alpha_0) + \delta_a^{A'}(\alpha_0, \epsilon, m_V)
\right),
\label{eq:Ea_true}
\end{eqnarray}
where $n$ denotes the principal quantum number, and the subindex, $a$, is used to distinguish the atomic states.
In the parenthesis, the first term is the leading contribution, which is obtained by solving the Schr\"odinger equation of the hydrogen atom. The second term, $\delta_a^{\rm SM}(\alpha_0)$, is SM corrections, which include relativistic corrections, recoil corrections, self energies, vacuum polarizations, two- and three-photon (loop) corrections, and finite size corrections of the nuclei. They are known well, and the uncertainties are controlled at the level of ${\cal O}(10^{-12})$, where the leading error originates in the proton charge radius. See Ref.~\cite{Mohr:2012tt} for details of the theoretical evaluation. 
On the other hand, the hidden photon contributions are evaluated as
\begin{eqnarray}
\delta_a^{A'} = \frac{8}{n_a^3}\lrfp{\epsilon\alpha_0 m_e }{m_V}{2}
 \left( 1 + {\cal O}\left( \alpha_0 \right) \right) +{\cal O} \left( \epsilon ^4\right),
\label{eq:corr_Elevel}
\end{eqnarray}
where the leading term  is from a tree-level correction to the static potential. The hidden photon is exchanged between the electron and the nuclei. Since the correction is sufficiently small, this is evaluated by the first order perturbation theory of the hydrogen atom. Higher order corrections are negligible. 

Experimentally, the most precisely measured transition frequency is that of the $1S$--$2S$ levels,
$(E_{2S_{1/2}}-E_{1S_{1/2}})/h$, whose numerical result is found in Table~\ref{tab:numbers} of Appendix. 
This experimental value is compared to the theoretical prediction in Eq.~(\ref{eq:Ea_true}),
\begin{eqnarray}
\lrf{E_{2S_{1/2}}-E_{1S_{1/2}}}{hc}^{\rm exp} 
&=& 
R_{\infty}
\left(\frac{3}{4} + \delta_{1S_{1/2}}^{\rm SM} -  \delta_{2S_{1/2}}^{\rm SM}
+ \delta_{1S_{1/2}}^{A'} -  \delta_{2S_{1/2}}^{A'}\right).
\label{eq:Rydberg_darkphoton}
\end{eqnarray}
Note that the prediction includes the hidden photon contributions. 
If it is assumed that the corrections are only from the SM, 
the Rydberg constant is determined  by
\begin{eqnarray}
\lrf{E_{2S_{1/2}}-E_{1S_{1/2}}}{hc}^{\rm exp} 
&=& 
R_{\infty}^{\rm SM-hyp.}
\left(\frac{3}{4} + \delta_{1S_{1/2}}^{\rm SM} -  \delta_{2S_{1/2}}^{\rm SM}\right).
\label{eq:Rydberg_SM}
\end{eqnarray}
The numerical value in the literature~\cite{Mohr:2012tt} is based on the SM hypothesis. 
In the hidden photon model, $R_{\infty}^{\rm SM-hyp.}$ deviates from the Rydberg constant of the theory defined by Eq.~(\ref{eq:Rydberg}) as
\begin{eqnarray}
\frac{R_{\infty}^{\rm SM-hyp.}(\alpha_0, \epsilon, m_V)}{R_{\infty}}
&=&
1+\frac{4}{3}(\delta_{1S_{1/2}}^{A'} -  \delta_{2S_{1/2}}^{A'})(1+{\cal O}(\delta_a^{\rm SM}))
\nonumber \\
&\simeq &
1 + \frac{28}{3} \lrfp{\epsilon\alpha_0 m_e }{m_V}{2}.
\label{eq:Rydberg_measured}
\end{eqnarray}
This is a source of the hidden photon contribution to the fine structure constant, which is additional to the SM evaluation of the constant in the literature. 

\subsection{$\Ar{\rm e}$}
In this and next subsection we briefly review the measurements of the relative atomic masses, $\Ar{\rm e}$ and $\Ar{\Rb87}$, 
and discuss the possibility of hidden photon correction to the values. The relative atomic masses are measured by so-called the Penning trap method, where charged particles are trapped in the electric and magnetic fields.
Masses of the charged particles are determined by using equations of motion.
The relative atomic mass of a neutral atom, $\Ar{X} $, is related with that of the $n$-th ionized atom, $\Ar{X^{n+}}$, by
\begin{eqnarray}
\Ar{X} &=& \Ar{X^{n+}} + n \Ar{\rm e} - \frac{\Eb{X} - \Eb{X^{n+}}}{m_{\rm u} c^2},
\label{eq:XandXnplus}
\end{eqnarray}
where $\Eb{X}/m_{\rm u}c^2$ is the total binding energy scaled by $m_{\rm u}c^2 \equiv m_{\C12}c^2 / 12$, namely in units of the relative atomic mass. As for the neutral atom with the atomic number $Z$, the total binding energy is a sum of the binding energy of the $Z$ electrons, while it is that of the $(Z-n)$ electrons for the $n$-th ionized atom, $X^{n+}$~\cite{Mohr:2012tt}.

The most precise inferred value of $\Ar{\rm e}$ is found in Ref.~\cite{Mohr:2012tt} as
\begin{equation}
\Ar{\rm e} =5.485\ 799\ 0932\ (29) \times 10^{-4}\quad [5.2\times 10^{-10}].
\label{eq:Are}
\end{equation}
This is obtained from the $g$-factor of the electron bounded in a hydrogenic carbon ion, by using the following relation:
\begin{eqnarray}
\frac{f_s(\C12^{5+})}{f_c(\C12^{5+})}
=
- \frac{g_e(\C12^{5+})}{10} \frac{\Ar{\C12^{5+}}}{\Ar{\rm e}},
\label{eq:Are_eq}
\end{eqnarray}
where $f_s$ and $f_c$ are the electron's precession frequency and the cyclotron frequency of the ion, respectively. 
The ratio $f_s/f_c(\C12^{5+})$ is precisely measured~\cite{fsfc}, which is listed in Table~\ref{tab:numbers}.
On the other hand, $\Ar{\C12^{5+}}$ is derived from Eq.~(\ref{eq:XandXnplus}) with $\Ar{\C12}=12$ by definition, 
where the relative binding energy is known with high precision by measurements of the carbon spectroscopy (see Table~\ref{tab:numbers}).
The theoretical value of the $g$-factor is known as $g_e(\C12^{5+})=-2.001\ 041\ 590\ 181\ (26)\; [1.3\times 10^{-11}]$~\cite{Mohr:2012tt}. With these numbers, one can reproduce the number in Eq.~(\ref{eq:Are}).

Let us examine the hidden photon effects. First of all, the relation \eqref{eq:XandXnplus} is independent of the hidden photon corrections as long as the binding energies are determined by experiments. Next, Eq.~\eqref{eq:Are_eq} is derived by equations of motions of the cyclotron and the spin precession in the magnetic field of the Penning trap. Since the experiment is macroscopic, the hidden photon correction is negligible. However, the input parameter, $g_e(\C12^{5+})$, is set theoretically in Eq.~\eqref{eq:Are_eq} and can be affected by the hidden photon. The leading contribution is the same as $g_e$ in vacuum,
\begin{eqnarray}
\Delta g_e^{A'} (\C12^{5+})  =2\Delta a_e ^{A'} +{\cal O} (\alpha_0).
\label{eq:corr_g_e}
\end{eqnarray}
The difference between $\Delta g_e^{A'}(\C12^{5+})$ and $\Delta g_e^{A'}$ in vacuum is from the Coulomb potential, which is of ${\cal O}(\alpha _0)$ and unimportant here.
Thus, from Eqs.~(\ref{eq:lepton_g-2}), (\ref{eq:Are_eq}) and (\ref{eq:corr_g_e}), compared to the SM hypothesis, the determination of $\Ar{\rm e}$ is affected by the hidden photon as
\begin{eqnarray}
\frac{\Ar{\rm e}^{\rm SM-hyp.}(\alpha_0, \epsilon, m_V)}{\Ar{\rm e}}
 &\simeq& 
 1 - \frac{\alpha_0}{3\pi } \lrfp{\epsilon m_e }{m_V}{2}.
\label{eq:Are_corr}
\end{eqnarray}

\subsection{$\Ar{\Rb87}$}
Similarly to $\Ar{\rm e}$ in the last subsection, $\Ar{\Rb87}$ is precisely measured by the Penning trap.
Circulating a couple of ions in the same magnetic field, a ratio of the relative atomic masses is determined by measuring a cyclotron frequency ratio of the ions. From the equations of motions in the magnetic field, the relation is
\begin{eqnarray}
\frac{f_c({\rm Y}^{m+})}{f_c({\rm X}^{n+})}
&=&
\frac{m}{n}\cdot 
 \frac{\Ar{X^{n+}}}{\Ar{Y^{m+}}}.
\end{eqnarray}
In particular, $f_c(\Rb87^{2+})/f_c(\Kr86^{2+})$ and 
$f_c(\C12\oo16_2^+)/f_c(\Kr86^{2+})$ have been measured
very precisely~\cite{PRA82:042513,PRA79:012506}, 
as listed in Table~\ref{tab:numbers}. Thus, 
$\Ar{\Rb87^{2+}}$ is related to $\Ar{\C12\oo16_2^+}$ by
\begin{eqnarray}
\Ar{\Rb87^{2+}} = 2
\lrf{f_c(\Rb87^{2+})}{f_c(\Kr86^{2+})}
\lrf{f_c(\Kr86^{2+})}{f_c(\C12\oo16_2^+)}
\Ar{\C12\oo16_2^+}.
\end{eqnarray}
The ion masses, $\Ar{\Rb87^{2+}}$ and $\Ar{\C12\oo16_2^+}$, are related to
the neutral ones, $\Ar{\Rb87}$ and $\Ar{\C12\oo16_2}$,
respectively, by Eq.~(\ref{eq:XandXnplus}), in which the relative binding energies 
have been experimentally measured, as listed in Table~\ref{tab:numbers}.
On the other hand, $\Ar{\C12\oo16_2}$ is obtained by
\begin{equation}
 \Ar{\C12\oo16_2} =\Ar{\C12} +2\Ar{\oo16} +{\rm bond\ energy},
\end{equation}
where the bond energy of CO$_2$ is also known (see Table~\ref{tab:numbers}).
Finally, $\Ar{\oo16}$ is obtained by Eq.~(\ref{eq:XandXnplus}) and 
precisely measured values of the cyclotron frequency ratios of
$f_c(\oo16^{6+})/f_c(\C12^{4+})$
and $f_c(\oo16^{6+})/f_c(\C12^{6+})$~\cite{IJMS251:231}, which are summarized in Table~\ref{tab:numbers}.
Combining all the numbers together with $\Ar{\rm e}$ in Eq.~(\ref{eq:Are}),
we obtain $\Ar{\Rb87} = 86.909\ 180\ 541\ (13)$,
which is consistent with the referred value in \cite{PRA82:042513}, $\Ar{\Rb87} = 86.909\ 180\ 535\ (10)\; [1.2\times 10^{-10}]$.

In the above derivation, since the equations of motions are macroscopic, and the binding/bond energies are measured by the experiments, the hidden photon contributions are negligible. Although the hidden photon can contribute to $\Ar{\rm e}$ as discussed in the last subsection, the correction to $\Ar{\Rb87}$ is negligibly small, as estimated to be
\begin{eqnarray}
\frac{\Ar{\Rb87}^{\rm SM-hyp.}(\alpha_0, \epsilon, m_V)}{\Ar{\Rb87}}
&\simeq& 1 + \frac{\alpha_0}{3\pi } \lrfp{\epsilon m_e }{m_V}{2} {\cal O} \left(  \frac{\Ar{\rm e}}{\Ar{\Rb87}}\right) .
\label{eq:ArRb_corr}
\end{eqnarray}


\subsection{$h / m_{\rm Rb} $}
The last piece of Eq.~(\ref{eq:alpha_exp}) is $h / m_{\rm Rb}$, which is a ratio of the Plank constant to the mass of the rubidium.
The measurement of the rubidium mass, $m_{\rm Rb}$, is responsible for the largest uncertainty of the fine structure constant, i.e., the theoretical prediction of the electron $g-2$. At present, the most precise result is based on an ingenious experiment using atomic interferometer~\cite{FineStructureConstant}. The basic idea is to measure the kinetic momentum/energy of the rubidium atom. The rubidium atoms are accelerated by the absorptions and the induced--emissions of photons, or by so-called the Bloch oscillation. Then, the shift of the kinetic energy changes a frequency of the wave function of the rubidium atom. The phase shift due to the acceleration is measured by the Ramsey-Bord${\rm \acute{e}}$ atomic interferometer: before the acceleration, the rubidium atoms are split into two packets by using the laser beams which are tuned to form so-called the $\pi/2$ pulse. By propagating these packets differently, the acceleration induces a difference of the phase of the wave functions. The relative phase, i.e., the kinetic energy correction from the acceleration, is measured by making the two packets interfere with each other after merging them by hitting the $\pi/2$ pulses again. From the conservations of the energy and momentum, the relative frequency (phase) of the two packets, $\Delta\omega$, is related to $h / m_{\rm Rb}$ as 
\begin{equation}
2\pi |\Delta\omega| = 2Nk_B (k_1+k_2) \frac{h}{m_{\rm Rb}},
\label{eq:h/m}
\end{equation}
where $2 N k_B$ is a total contribution to the wave number of the rubidium packets by the Bloch oscillations, and $k_1$ and $k_2$ are wave numbers of the latter $\pi/2$ pulse, which consists of the two laser beams with the wave number, $k_1$ and $k_2$, in the opposite direction. From \eqref{eq:h/m}, the interference pattern of the atomic interferometer is obtained by sweeping $k_1$ and $k_2$. Since the frequencies of the laser beams are known very precisely, and the interference pattern is sensitive to the relative phase, $h / m_{\rm Rb}$ is measured with high precision as
\begin{equation}
\frac{h}{m_{\rm Rb}}=4.591\ 359\ 2729(57)\times 10^{-9} ~{\rm m^2s^{-1}}\quad [1.2\times 10^{-9}].
\end{equation}

The measurement of the rubidium mass is insensitive to the hidden photon effects. The relation (\ref{eq:h/m}) is based only on the energy and momentum conservations, and the all the quantities except for $h / m_{\rm Rb}$ are given in the experiment. Although the precision of the Ramsey-Bord${\rm \acute{e}}$ atomic interferometer relies on the (meta-) stability of the atoms during the propagation against spontaneous emissions, the hidden photon in interest is too weak to enhance the emission and to spoil the experiment. Therefore, the measurement of $h / m_{\rm Rb}$ in the hidden photon model coincides with that of the SM hypothesis as
\begin{eqnarray}
\frac{(h/m_{\rm Rb})^{\rm SM-hyp.}(\alpha_0, \epsilon, m_V)} {h/m_{\rm Rb}} 
&=&1.
\label{eq:Rb_corr}
\end{eqnarray}

\subsection{Summary of hidden photon contributions}

From Eqs.~(\ref{eq:alpha_exp}), (\ref{eq:Rydberg_measured}),
(\ref{eq:Are_corr}), (\ref{eq:ArRb_corr}), and (\ref{eq:Rb_corr}), 
the contribution of the hidden photon to the fine structure constant becomes 
\begin{eqnarray}
\frac{\alpha^{\rm SM-hyp.}(\alpha_0, \epsilon, m_V)}{\alpha_0}
&\simeq& 1 + \left(\frac{14\alpha_0^2}{3} - \frac{\alpha_0}{6\pi}\right)
\lrfp{\epsilon m_e }{m_V}{2},
\label{eq:correction_to_alpha}
\end{eqnarray}
which leads to
\begin{eqnarray}
\Delta a_e (\alpha_0, \epsilon, m_V)^{{\rm from}\;\alpha} 
&\simeq& -\frac{\alpha_0}{2\pi}
\left(\frac{14\alpha_0^2}{3} - \frac{\alpha_0}{6\pi}\right)
\lrfp{\epsilon m_e }{m_V}{2}\\
&\simeq&
-\left(7\alpha_0^2 - \frac{\alpha_0}{4\pi}\right) \Delta a_e^{A'{\rm loop}}
\label{eq:from_alpha}
\end{eqnarray}
where Eq.~(\ref{eq:lepton_g-2}) was used in the second equality. 

\section{Constraint from electron $g-2$}

The electron $g-2$ receives corrections from Eqs.~\eqref{eq:lepton_g-2} and \eqref{eq:from_alpha}. 
\begin{equation}
a_e(\alpha_0 )=a_e ^{\rm SM} (\alpha^{\rm SM-hyp.})+\Delta a_e^{A'{\rm loop}} (\alpha_0)+\Delta a_e^{{\rm from}\;\alpha} (\alpha_0).
\end{equation}
The correction through the determination of the fine structure constant is negligibly small compared to the loop contribution, and the dominant contribution is from the one-loop correction \eqref{eq:lepton_g-2}.

\begin{figure}[t]
\begin{center}
\includegraphics[scale=0.6]{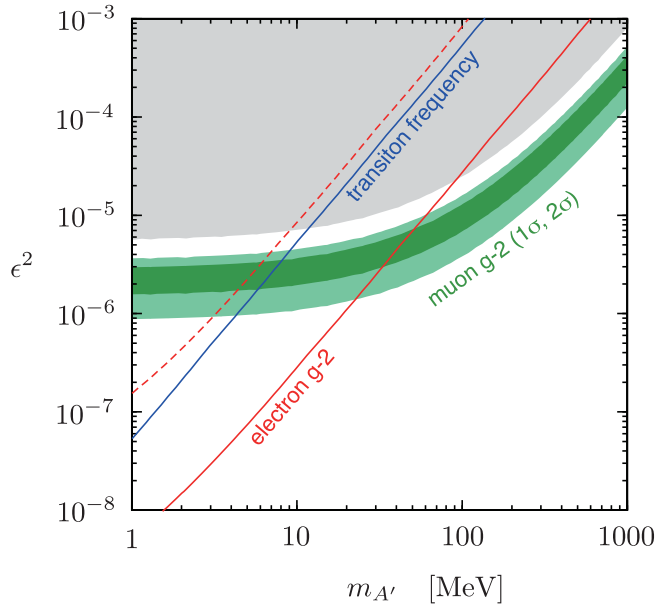}
\caption{The constraints on the hidden photon model are shown. The red--solid line is updated by the analysis in this paper, which is compared to the previous one~\cite{Pospelov:2008zw}, drawn by the red--dashed line. The constraint from the transition frequencies is also displayed by the blue solid line, where the $1S_{1/2}-2S_{1/2}$ and $2S_{1/2}-8D_{5/2}$ transitions are analyzed with the proton radius input by the electron--proton experiment~\cite{Mohr:2012tt,Bernauer:2010wm}. The regions favored by the muon $g-2$ at the 1 and 2$\sigma$ levels are shown by the dark and light green bands, respectively.}
\label{fig:darkphoton}
\end{center}
\end{figure}

\begin{figure}[t]
\begin{center}
\includegraphics[scale=0.4]{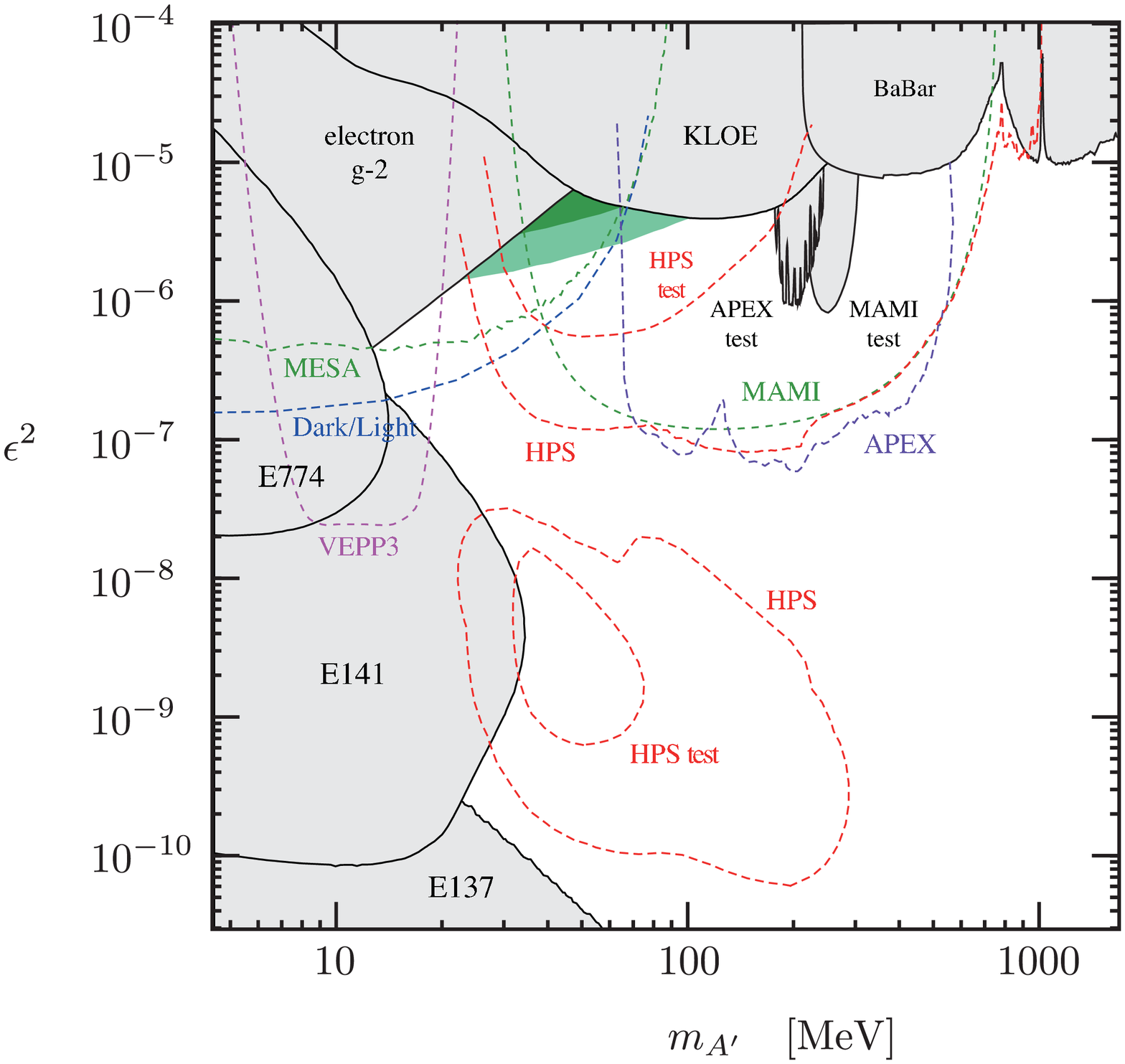}
\caption{The updated constraint from the electron $g-2$ is compared to the other experimental bounds~\cite{Hewett:2012ns,Babusci:2012cr}. The gray regions have been already excluded by them. The green bands are favored by the muon $g-2$ (see Fig.~\ref{fig:darkphoton}). The regions enclosed by the colored--dashed lines are suggested to be covered in future. See Ref.~\cite{Hewett:2012ns} for details of the experiments, where the excluded regions and future sensitivities are found. The recent update of the KLOE bound~\cite{Babusci:2012cr} is included (see Note Added).}
\label{fig:combined}
\end{center}
\end{figure}

In Fig.~\ref{fig:darkphoton}, the constraints on the hidden photon model are displayed. The red--solid line is the updated result of the bound from the electron $g-2$, which excludes the region above the line at the 2$\sigma$ level. This result is improved by one order of magnitude compared to the previous analysis, which is shown by the red--dashed line in the figure~\cite{Pospelov:2008zw}. It is interesting to compare the result with the region favored by the muon $g-2$. The 1 and 2$\sigma$ regions are shown by the dark and light green bands, respectively, in the figure. The electron $g-2$ constraint now excludes a smaller mass region of the hidden photon more efficiently. In Fig.~\ref{fig:combined}, the updated constraint is compared to the other experimental bounds~\cite{Hewett:2012ns}. Before the present analysis, the lower mass region has been excluded mainly by the beam dump experiment at E774~\cite{Bross:1989mp}. As shown in the figure, the region which is favored by the muon $g-2$ becomes rejected by the electron $g-2$. It is found that  the allowed region of the muon $g-2$ can be accessed by various experiments, which are displayed by the dashed lines in Fig.~\ref{fig:combined}~\cite{Hewett:2012ns}. Interestingly, even the test run of the HPS experiment, which is planned in the near future~\cite{PATRAS2012-talk}, can cover most of the parameter space. 

\section{Hydrogen transition frequencies}
\label{sec:new_constraint}

Among the atomic parameters, the transition frequencies are measured with extremely high precision. In fact, the accuracy is better than that of the electron $g-2$.
The transition frequencies can provide search for the hidden photon, which is independent of the electron $g-2$.
  In order to compare the experimental result with the theoretical prediction, measurements of the input parameters should be independent of or at least insensitive to that of the transition frequencies. In this section, the transition frequencies are studied to search for the hidden photon in detail. 

A transition frequency is written by the Rydberg constant and the corrections from the SM and the hidden photon as 
\begin{equation}
  \frac{E_a - E_b}{h} =  - R_{\infty}c 
  \left[ 
  \left( \frac{1}{n_a^2} - \frac{1}{n_b^2} \right) 
  + \left( \delta_a^{\rm SM} - \delta_b^{\rm SM} \right)
  + \left( \delta_a^{A'} - \delta_b^{A'} \right)
  \right],
  \label{eq:transition}
\end{equation}
where the left-handed side is measured by experiments, while the quantities in the bracket are theoretically given. In particular, the SM corrections, $\delta^{\rm SM}$, are well known for the hydrogen atom. They consist of relativistic corrections, recoil corrections, self energies, vacuum polarizations, two- and three-photon (loop) corrections, and finite size corrections of the proton. The detailed evaluations are summarized in \cite{Mohr:2012tt}. The relative uncertainty of the prediction is of ${\cal O}(10^{-12})$. The leading part comes from the measurement of the proton charge radius, which will be mentioned later, and the next-to-leading one, which is of ${\cal O}(10^{-13})$, is from ambiguity of the evaluation of the two-photon correction (see discussions in \cite{Mohr:2012tt}). On the other hand, the hidden photon shifts the energy levels by $\delta^{A'}$, which is obtained in Eq.~\eqref{eq:corr_Elevel}. As we will discuss later, the other contributions are negligible. 

Since the correction from the hidden photon $\delta^{A'}$ is tiny, if Eq.\eqref{eq:transition} is used to constrain the hidden photon model, it is crucial to determine the Rydberg constant $R_{\infty}$ with extremely high precision. Although $R_{\infty}$ has been provided very precisely in CODATA~\cite{Mohr:2012tt}, we should not refer to it in the current analysis. This is because the determination of the Rydberg constant 
is based on the transition frequencies, and hence the analysis would become self--inconsistent.\footnote{
The treatment of $R_{\infty}$ is unclear in \cite{Jaeckel:2010xx}, in which they refer to the CODATA, despite that the hidden photon contributions to the transition frequencies are studied. 
}
On the other hand, if the Rydberg constant is determined from the definition, \eqref{eq:Rydberg}, it is required to know the fine structure constant and the electron mass very precisely. If the lepton $(g-2)$'s are used to determine the fine structure constant, the hidden photon contribution to the transition frequency, $\delta^{A'}$, becomes obscure.  Also,  the relation \eqref{eq:alpha_exp} cannot be used to determine the $\alpha_0$, because it depends on $R_{\infty}$. Thus, other methods are required to determine $\alpha_0$ accurately. Furthermore, it is very difficult to measure the electron mass with the required precision. The accuracy of $\Ar{\rm e}$ is worse than that of Eq.~\eqref{eq:transition}.

In order to avoid the difficulties of the Rydberg constant, let us consider a ratio of the transition frequencies,
\begin{eqnarray}
  \label{eq:ratio_transition}
  \frac{(E_a - E_b)/h}{(E_c - E_d)/h} &=&  
  \frac
  {
  \left( \frac{1}{n_a^2} - \frac{1}{n_b^2} \right) 
  + \left( \delta_a^{\rm SM} - \delta_b^{\rm SM} \right)
  + \left( \delta_a^{A'} - \delta_b^{A'} \right)
  }
  {
  \left( \frac{1}{n_c^2} - \frac{1}{n_d^2} \right) 
  + \left( \delta_c^{\rm SM} - \delta_d^{\rm SM} \right)
  + \left( \delta_c^{A'} - \delta_d^{A'} \right)
  } \\
  &\simeq&
  \left. \frac{(E_a - E_b)/h}{(E_c - E_d)/h} \right|_{\rm SM}
  + \frac{n_a^2 n_b^2}{n_b^2 - n_a^2} \left( \delta_a^{A'} - \delta_b^{A'} \right)
  - \frac{n_c^2 n_d^2}{n_d^2 - n_c^2} \left( \delta_c^{A'} - \delta_d^{A'} \right),
  \nonumber 
\end{eqnarray}
where the left-handed side is determined by experiments, and 
$(E_a - E_b)/h|_{\rm SM}$ is defined by the limit of $\delta^{A'} \to 0$,
\begin{eqnarray}
  \left. \frac{(E_a - E_b)/h}{(E_c - E_d)/h} \right|_{\rm SM} = 
  \frac
  {
  \left( \frac{1}{n_a^2} - \frac{1}{n_b^2} \right) 
  + \left( \delta_a^{\rm SM} - \delta_b^{\rm SM} \right)
  }
  {
  \left( \frac{1}{n_c^2} - \frac{1}{n_d^2} \right) 
  + \left( \delta_c^{\rm SM} - \delta_d^{\rm SM} \right)
  }.
\end{eqnarray}
It is noted that the Rydberg constant disappears. 

The SM correction, $\delta^{\rm SM}$, is precisely known~\cite{Mohr:2012tt}. It is a function of $\alpha$, $m_\mu/m_e$, $m_p/m_e$, and the proton root-mean-square charge radius, $r_p$ scaled by the Compton wavelength of electron, $\lambda_C = h/m_e c$. Among them, hidden photon contributions through $m_\mu/m_e$, $m_p/m_e$ and $\lambda_C$ can be safely neglected, since the corrections $\delta^{\rm SM}$ are less sensitive to these parameters.  
On the other hand, in the following analysis, the fine structure constant is determined by the rubidium atom based on the previous section. The contribution of the hidden photon gives arise mainly through the electron $g-2$ in the carbon atom \eqref{eq:correction_to_alpha}. However, as we will be see later, it is orders of magnitude smaller than the contribution in the energy level shift, $\delta^{A'}$. 

The proton charge radius is currently determined by the transition frequencies as well as the electron--proton scattering experiment~\cite{Mohr:2012tt}. 
In our analysis, 
the recommended value by CODATA~\cite{Mohr:2012tt} cannot be applied, since
$r_p$ is mainly determined by the spectroscopic data there.
We will instead use the result from the electron--proton scattering experiment, $r_p = 0.8971(79)\,{\rm fm}$~\cite{Mohr:2012tt,Bernauer:2010wm}. Its determination is considered to be insensitive to the hidden photon effects. Since the scattering cross section is dominated by the QED as $\sigma(ep \to ep) \propto \alpha^2$, the hidden photon contributions are at most $\mathcal{O}(\epsilon^2)$. Thus, in the parameter region where the muon $g-2$ is explained, the effect is blinded behind the experimental uncertainties. 
On the other hand, a more accurate proton radius has been obtained by measuring the Lamb shift of the muon hydrogen between $2S_{1/2}-2P_{3/2}$~\cite{Pohl:2010zza}. Unfortunately, the result is inconsistent with the CODATA recommended value as well as the electron--proton scattering result. This problem may be solved by corrections to the SM estimation (see, for instance, \cite{protonradius} for a review). In the following analysis, we adopt the electron--proton scattering result by following the argument by \cite{Mohr:2012tt}. 

Among the transition frequencies, the levels between $1S_{1/2} - 2S_{1/2}$ of the hydrogen has been measured with the best precision (see Table~\ref{tab:numbers})~\cite{Mohr:2012tt,PRL107_203001}. 
It turns out that the severest constraint on $\delta^{A'}$ is obtained from the ratio of the transition frequencies of $1S_{1/2} - 2S_{1/2}$ and $2S_{1/2} - 8D_{5/2}$~\cite{Mohr:2012tt}.
The relation \eqref{eq:ratio_transition} becomes
\begin{equation}
\left. \frac{(E_{2S_{1/2}}-E_{8D_{5/2}})/h}{(E_{1S_{1/2}}-E_{2S_{1/2}})/h} \right|_{\rm exp} -
\left. \frac{(E_{2S_{1/2}}-E_{8D_{5/2}})/h}{(E_{1S_{1/2}}-E_{2S_{1/2}})/h} \right|_{\rm SM} 
\simeq 
- \frac{4}{3} \delta_{1S_{1/2}}^{A'} + \frac{28}{5} \delta_{2S_{1/2}}^{A'} - \frac{64}{15} \delta_{8S/D}^{A'},
\label{eq:thVSexp}
\end{equation}
where the SM corrections are calculated by following the analysis in Ref.~\cite{Mohr:2012tt}. From the experimental data and the theoretical calculations with the input parameters discussed above, we obtain
\begin{equation}
\left. \frac{(E_{2S_{1/2}}-E_{8D_{5/2}})/h}{(E_{1S_{1/2}}-E_{2S_{1/2}})/h} \right|_{\rm exp} -
\left. \frac{(E_{2S_{1/2}}-E_{8D_{5/2}})/h}{(E_{1S_{1/2}}-E_{2S_{1/2}})/h} \right|_{\rm SM} 
= (1.9 \pm 2.9) \times 10^{-12},
\label{eq:bound-freq-num}
\end{equation}
where the largest uncertainty is from the experimental result of $2S_{1/2} - 8D_{5/2}$, while the theoretical error, which originates in the proton radius, is half of the experimental one. 
Therefore,  the constraint is obtained as
\begin{equation}
  \frac{77}{15} \lrfp{\epsilon\alpha_0 m_e }{m_V}{2} < 3.9 \times 10^{-12},
  \label{eq:bound-freq}
\end{equation}
at the 2$\sigma$ level. We have checked that if the hidden photon contribution to the fine structure constant is taken into account, the numerical factor, $77/15$, is corrected by $\mathcal{O}(10^{-5})$, and thus, it is negligible. 

In Fig.~\ref{fig:darkphoton}, the constraint from the transition frequency is plotted. Above the blue--solid line, the hidden photon model is excluded by the hydrogen spectroscopy. The current limit is weaker by about one order of magnitude than that of  the electron $g-2$. However, the constraint is independent of the electron $g-2$, and the sensitivity is expected to be improved in future. 
The uncertainty of Eq.~\eqref{eq:bound-freq-num} is dominated by the experimental measurement of $(E_{2S_{1/2}}-E_{8D_{5/2}})/h$. It is expected to be reduced, since the experiment is in progress at the National Physics Laboratory~\cite{exp_2SnSD}. In addition, the theoretical error is mainly from the proton charge radius. If the inconsistency between the values of the electron--proton scattering and the muon hydrogen will be resolved, the uncertainty in the latter result is smaller by one order of magnitude than that of the former, and the theoretical error associated to the proton radius will decrease greatly. The sub-leading contribution to the theoretical error is from the two-photon correction to the SM prediction. Since this is of ${\cal O}(10^{-13})$, the constraint from the transition frequency can be comparable to that of the electron $g-2$ in future. 


\section{Conclusion}
The hidden photon model is one of the simplest models which can explain the muon $g-2$ anomaly. In this paper, we have examined the constraints on the hidden photon model from the electron $g-2$ as well as from the precise measurements of atomic parameters. We have shown that the new constraint from the electron $g-2$ is more than one order of magnitude stronger than the previous analysis. The parameter space which can explain the muon $g-2$ anomaly becomes narrow, $m_V\simeq 20$--100 MeV and $\epsilon^2\simeq 10^{-6}$--$10^{-5}$. Interestingly, this region can be covered by experiments in future~\cite{Hewett:2012ns}.

The theoretical prediction of the electron $g-2$ is based on a precise measurement of the fine structure constant, which is determined independently of the $g-2$. We have also discussed the effect of the hidden photon on the determination of the fine structure constant. We have shown that, although the determination of the fine structure constant is affected by the hidden photon, its contribution to the electron $g-2$ is small compared with the direct loop contribution to the $g-2$.

We have also investigated a constraint on the hidden photon model from the precise measurements of hydrogen transition frequencies. Since the atomic energy levels are affected by the hidden photon, the precisely measured transition frequencies can be used to constrain the hidden photon model. A careful analysis is essential, since the determination of the various quantities, such as the Rydberg constant or transition frequencies, are often correlated. We have shown that, by taking a ratio of transition frequencies, the measurement becomes free from the uncertainty of the Rydberg constant, and a severe constraint is obtained. The constraint on the hidden photon model is now about one order of magnitude weaker than the electron $g-2$, but more precise measurements of the transition frequencies can improve the constraint, which may compete with the $g-2$ constraint.

\section*{Note Added}
While completing this paper, we noticed that the version 2 of Ref.~\cite{Davoudiasl:2012he} was posted on arXiv, which also updated the bound on the hidden photon model from the electron $g-2$. After this paper was posted, the KLOE experiment updated the constraint at the 90\% C.L.~\cite{Babusci:2012cr}, which is shown in Fig.~\ref{fig:combined}.

\section*{Acknowledgments}
This work was supported by Grand-in-Aid for Scientific research from 
the Ministry of Education, Science, Sports, and Culture (MEXT), Japan,
No. 23740172 (M.E.), No. 21740164 (K.H.), and No. 22244021 (K.H.).
This work was supported by World Premier International Research Center Initiative (WPI Initiative), MEXT, Japan.

\appendix
\section{Experimental values}
In this appendix, we list some values of precisely known quantities, which are used in Sec.~\ref{sec:alpha} and Sec.~\ref{sec:new_constraint}.
In Table~\ref{tab:numbers}, the hydrogen transition frequencies are shown for $2S_{1/2}$--$1S_{1/2}$ and $8D_{5/2}$--$2S_{1/2}$ transitions. The effects of hyper--fine splitting is corrected in the table. The hidden photon affects the hyper--fine splitting through the magnetic moments, but its  contribution is smaller than the experimental uncertainties. 
 The ratios of cyclotron frequencies (and electron's precession frequency) are also shown for the relevant ions.

In the table, the relative binding energies of atoms and ions are also listed. 
We could not find the uncertainties for some of the binding energies, but we expect these uncertainties are close to the ones next to them.
Binding energies are measured in terms of cm$^{-1}$~\cite{energy-C, energy-O}, and
conversion into units of eV or relative atomic mass unit may generate uncertainty.
However, uncertainties coming from these conversion factors are negligibly small.

\renewcommand{\arraystretch}{1.4}
\begin{table}
\begin{center}
\begin{tabular}{r @{} lcr @{} l@{}lll}
\hline
&& && && relative &
\\
\multicolumn{2}{c}{quantity} & \multicolumn{3}{c}{measured value} &  & uncertainty & Ref.
\\ \hline
$(E_{2S_{1/2}}$&  $-E_{1S_{1/2}})/h$
&\multicolumn{3}{l}{
2\ 466\ 061\ 413\ 187.035\ (10) }& kHz 
&
[$4.2\times 10^{-15}$]
& 
\cite{PRL107_203001}
\\
$(E_{8D_{5/2}}$&$-E_{2S_{1/2}})/h$
&\multicolumn{3}{l}{
\;\; 770\ 649\ 561\ 584.2\ (6.4) }& kHz
&
[$8.3\times 10^{-12}$]
& 
\cite{Mohr:2012tt}
\\ \hline
$f_s(\C12^{5+})$&$/f_c(\C12^{5+})$&
&\multicolumn{3}{c}
{
4376.210\ 4989\ (23) }&

[$5.3\times 10^{-10}$]
&
\cite{fsfc}
\\ 
$f_c(\Rb87^{2+})$&$/f_c(\Kr86^{2+})$&&
0.988&\multicolumn{2}{l}{510\ 045\ 784\ (69)} &

[$7.0\times 10^{-11}$]
&
\cite{PRA82:042513}
\\ 
$f_c(\C12\oo16_2^+)$&$/f_c(\Kr86^{2+})$&&
0.976
&\multicolumn{2}{l}{
482\ 363\ 817\ (128)} &
 
[$1.3\times 10^{-10}$]
&
\cite{PRA79:012506}
\\ 
$f_c(\oo16^{6+})$&$/f_c(\C12^{4+})$&&
1.125
&\multicolumn{2}{l}{
383\ 463\ 469\ (24)} &

[$2.1\times 10^{-11}$]
&
\cite{IJMS251:231}
\\
$f_c(\oo16^{6+})$&$/f_c(\C12^{6+})$&&
0.750
&\multicolumn{2}{l}{
187\ 093\ 102\ (8) }&

[$1.1\times 10^{-11}$]
&
\cite{IJMS251:231}
\\ \hline
$\Eb{\C12}$&$-\Eb{\C12^{+}}$&
&\quad\quad
\ 11&.260 296\ (12) & eV
&
[$1.1\times 10^{-6}$]
&
\cite{energy-C}
\\
$\Eb{\C12^{+}}$&$-\Eb{\C12^{2+}}$&
&
\ 24&.383 314 & eV
&
&
\cite{energy-C}
\\
$\Eb{\C12^{2+}}$&$-\Eb{\C12^{3+}}$&
&
\ 47&.887 77\ (25) & eV
&
[$5.2\times 10^{-6}$]
&
\cite{energy-C}
\\$\Eb{\C12^{3+}}$&$-\Eb{\C12^{4+}}$&
&
\ 64&.493 89\ (19) & eV
&
[$2.9\times 10^{-6}$]
&
\cite{energy-C}
\\$\Eb{\C12^{4+}}$&$-\Eb{\C12^{5+}}$&
&
392&.0869\ (37) & eV
&
[$9.5\times 10^{-6}$]
&
\cite{energy-C}
\\ \hline
$\Eb{\oo16}$&$-\Eb{\oo16^{+}}$&
&
\ 13&.618 0543\ (74) & eV
&
[$5.5\times 10^{-7}$]
&
\cite{energy-O}
\\
$\Eb{\oo16^{+}}$&$-\Eb{\oo16^{2+}}$&
&
\ 37&.634 6826 & eV
&
&
\cite{energy-O}
\\
$\Eb{\oo16^{2+}}$&$-\Eb{\oo16^{3+}}$&
&
\ 54&.935 49\ (24) & eV
&
[$4.5\times 10^{-6}$]
&
\cite{energy-O}
\\$\Eb{\oo16^{3+}}$&$-\Eb{\oo16^{4+}}$&
&
\ 77&.413 49 & eV
&
&
\cite{energy-O}
\\$\Eb{\oo16^{4+}}$&$-\Eb{\oo16^{5+}}$&
&
113&.899 07\ (50) & eV
&
[$4.4\times 10^{-6}$]
&
\cite{energy-O}
\\
$\Eb{\oo16^{5+}}$&$-\Eb{\oo16^{6+}}$&
&
138&.1196 & eV
&
&
\cite{energy-O}
\\$\Eb{\oo16^{6+}}$&$-\Eb{\oo16^{7+}}$&
&
739&.292\ (37) & eV
&
[$5.0\times 10^{-5}$]
&
\cite{energy-O}
\\\hline
$\Eb{\Rb87}$&$-\Eb{\Rb87^+}$&
&
\ \ 4&.117 128\ (1) & eV
&
[$2.4\times 10^{-7}$]
&
\cite{energy-Rb}
\\
$\Eb{\Rb87^+}$&$-\Eb{\Rb87^{2+}}$&
&
\ 27&.2898\ (1) & eV
&
[$3.7\times 10^{-6}$]
&
\cite{energy-Rb}
\\\hline
$\Eb{\rm CO_2}$&$-\Eb{\rm CO_2^+}$ &
&
\ 13&.773\ (2) & eV
&
[$1.5\times 10^{-4}$]
&\cite{energy-CO2}
\\
$\Delta E({\rm CO_2}$&$\to{\rm CO+O})$&
&
\ \ 5&.451\ (4) & eV
& [$7.6\times 10^{-4}$]
&\cite{bond_energy}
\\
$\Delta E({\rm CO}$&$\to{\rm C+O})$&
&
\ 11&.110\ (4) & eV
&[$3.7\times 10^{-4}$]
&\cite{bond_energy}
\\\hline
\end{tabular}
\caption{Experimentally measured quantities which are used in the discussion in Sec.~\ref{sec:alpha} and Sec.~\ref{sec:new_constraint}. In the last two lines, $\Delta E$ denotes the bond energies.}
\label{tab:numbers}
\end{center}
\end{table}

  
\end{document}